\documentclass[9pt, conference]{IEEEtran}
\IEEEoverridecommandlockouts

\usepackage{cite}
\usepackage{amsmath,amssymb,amsfonts}
\usepackage{algorithmic}
\usepackage{url}
\usepackage{multirow}
\usepackage{graphicx}
\usepackage{textcomp}
\usepackage{xcolor}
\usepackage{subcaption} 

\def\BibTeX{{\rm B\kern-.05em{\sc i\kern-.025em b}\kern-.08em
    T\kern-.1667em\lower.7ex\hbox{E}\kern-.125emX}}
\begin{document}

\title{Explaining Speaker and Spoof Embeddings via Probing
\thanks{This study was carried out using the TSUBAME4.0 supercomputer at Institute of Science Tokyo. We specially thank Xin Wang for his fruitful suggestions along the discussion on the idea and experiments.}
}

\author{\IEEEauthorblockN{Xuechen Liu\IEEEauthorrefmark{1}, Junichi Yamagishi\IEEEauthorrefmark{1}, Md Sahidullah\IEEEauthorrefmark{2}, Tomi Kinnunen\IEEEauthorrefmark{3}}
    \IEEEauthorblockA{\IEEEauthorrefmark{1}National Institute of Informatics, 2-1-2 Hitotsubashi, 101-8430 Tokyo, Japan}
    \IEEEauthorblockA{\IEEEauthorrefmark{2}Institute for Advancing Intelligence, TCG CREST, 700791 Kolkata, India}
    \IEEEauthorblockA{\IEEEauthorrefmark{3}School of Computing, University of Eastern Finland, FI-80100 Joensuu, Finland}
\texttt{\{xuecliu, jyamagis\}@nii.ac.jp, md.sahidullah@tcgcrest.org, tomi.kinnunen@uef.fi}
}

\maketitle

\begin{abstract}
This study investigates the explainability of embedding representations, specifically those used in modern audio spoofing detection systems based on deep neural networks, known as spoof embeddings. Building on established work in speaker embedding explainability, we examine how well these spoof embeddings capture speaker-related information. We train simple neural classifiers using either speaker or spoof embeddings as input, with speaker-related attributes as target labels. These attributes are categorized into two groups: metadata-based traits (e.g., gender, age) and acoustic traits (e.g., fundamental frequency, speaking rate). Our experiments on the ASVspoof 2019 LA evaluation set demonstrate that spoof embeddings preserve several key traits, including gender, speaking rate, F0, and duration. Further analysis of gender and speaking rate indicates that the spoofing detector partially preserves these traits, potentially to ensure the decision process remains robust against them.
\end{abstract}

\section{Introduction}
\label{sec:intro}
Low-dimensional embedding vectors extracted using \emph{deep neural network} (DNN) models 
can be used to address different speech 
tasks, including \emph{text-to-speech} (TTS) \cite{dnn_speech_synthesis}, \emph{voice conversion} (VC) \cite{voice_conversion_review2020},  \emph{automatic speaker verification} (ASV) \cite{asv-dnn-review_2021}, and \emph{audio spoofing detection} \cite{logit_antispoofing2013}. Advances in the last have led to numerous \emph{countermeasure} (CM) algorithms for protecting ASV in the form of \emph{spoof-aware speaker verification} \cite{sasv2022}. 
Such spoofing detectors, which can be used to extract embeddings from their intermediate layers, have not only enhanced CM performance but also improved ASV system resilience against spoofing \cite{sasv_embedding2022, gen_sasv2024}.
Moreover, ASV and CM contains information that may complement each other. It has been demonstrated that the information from a well-trained CM module can can improve ASV robustness against spoofing attacks \cite{gen_sasv2024}. Likewise, the pre-enrolled speaker embedding vector can serve as auxiliary information to improve CM performance \cite{speaker_aware_cm2023}. Those findings raise an important question: \textbf{what kind of information is captured by speaker and spoof embeddings, and acquired for making their decisions}?

This study investigates the differences between ASV speaker embeddings and CM spoof embeddings to improve our understanding of spoof detection. Specifically, we aim to determine what elements are preserved in robust spoof detection systems and what information is discarded compared to standard speaker embeddings.
This is achieved by evaluating totally 10 \emph{traits} about speaker and spoof embeddings, covering both metadata from the source dataset, and acoustic properties of the speech. Hereafter, they are referred to as \emph{meta traits} and \emph{acoustic traits}, respectively. We perform \emph{probing} analysis \cite{probe_study1, probe_study2} on those traits individually, by designing close-set classification and regression tasks, and training simple DNN using partitioned data from the ASVspoof 2019 dataset \cite{asvspoof2019}. To our knowledge, this is the first comprehensive study on the explainability of audio spoofing embeddings through probing analysis. Our research is fundamentally different from earlier works on explainable neural CM \cite{explainable_cm1, explainable_cm2}: we put our sole focus on the explainability of representations from modern CMs.

Our research is motivated by two key insights from previous work. First, adding information as auxiliary vectors has been effective in enhancing various speech processing tasks, including speech recognition \cite{speaker_recognition_ivector2013}, speaker recognition \cite{deep_feature_asv2015}, and spoofing detection \cite{gen_sasv2024, speaker_aware_cm2023}. This success suggests that these vectors preserve information that can be valuable for tasks that requiring invariance to such information. Second, earlier studies have demonstrated that the information embedded in representations for specific tasks, such as speaker embeddings, can be explicitly probed and analyzed \cite{probe_xvectors, speaker_embedding_encode2017}. We aim to uncover how CM models recover or re-purpose information from speaker verification embeddings to counter and generalize against spoofing attacks, even when confronted with novel attack types.

\section{Probing embeddings}
\label{sec:counter_facts}
Similar to \cite{probe_xvectors} where x-vector \cite{xvector2018} speaker embeddings were found to contain information about several traits such as gender and speaking rate, we hypothesize that the embeddings trained on domain-specific data can correspond to moderate performance preserve the information corresponding to the traits introduced in this section. If this hypothesis holds, one should be able to train a predictor model to predict these traits, with the prediction performance being positively correlated with the amount of information preserved. This approach echoes earlier work on speaker embeddings \cite{probe_xvectors} and forms the basis of our empirical experiments. What makes this work unique is our analysis of both speaker and spoof embeddings in the context of anti-spoofing with domain-specific traits (such as type of attacking algorithms), where we explored the preservation of the traits through several related classification and regression tasks.

\subsection{Probing model}
We used \emph{multi-layer perceptron} (MLP) for our probing analysis. This architecture comprises an input linear layer, a hidden linear layer, and an output layer. Each layer, with the exception of the output layer, incorporates an affine transformation followed by a nonlinear activation function. The output dimension of the final layer varies depending on the category of target traits. For meta speaker traits, the output dimension corresponds to the number of available ground truth label values, as these are represented as one-hot encoding vectors. In this case, the network was trained using \emph{cross-entropy} (CE) loss. For acoustic traits, the final linear layer outputs a single scalar value, representing the estimation of the corresponding trait. In this instance, the network was trained using \emph{mean squared error} (MSE) loss, where the two inputs were the scores predicted and the label value, respectively. The generation of label values for regression is presented in section \ref{secsec:physicals}.

We employed Adam \cite{adam} throughout our experiments for model optimization, with a initial learning rate of 0.001 and step-wise decay schedule. The batch size was set to 32 and number of training epoch was 20. All experiments were conducted on CPU instances equipped with 40 Intel Xeon cores and 16GB of memory. The evaluation metric used in this study is detailed in Section \ref{secsec:evalaution}.

\subsection{Datasets: ASVspoof 2019 \& VCTK}
\label{secsec:datasets}
The dataset used in this study is the ASVspoof 2019 LA dataset \cite{asvspoof2019}, a widely adopted benchmark for ASV and CM against spoofing attacks. It includes 19 different attack algorithms derived from various TTS and VC systems. It is originated from the CSTR VCTK corpus \cite{vctk} (referred to as ``VCTK" hereafter), which provides the source (i.e., bonafide) waveform for generating spoofed utterances. The recordings were made in a hemi-anechoic chamber, ensuring high-quality audio. More interestingly to this study, VCTK contains rich meta resources, with not only transcripts of the speech but also detailed multiple metadata \emph{traits} about the speakers, including their speaker identity, age, and accent, while keeping their real identities anonymized. In the ASVspoof 2019 dataset, speaker identities were re-annotated, with a mapping provided for reference.

\subsection{Meta probing traits (for classification)}
\label{secsec:meta}
We consider two types of meta traits in the probing tasks. The first one is speaker-related, which can be derived from VCTK metadata. We extracted the speaker ID information from the mapping data and aligned it with the ASVspoof 2019 speaker IDs (beginning with “LA\_”). The traits are detailed below:
\begin{itemize}
    \item Speaker ID: The speaker IDs are mapped from VCTK to the ASVspoof dataset. The original VCTK speaker IDs are also anonymized, with no direct reference to real speaker identities. The ASVspoof 2019 LA dataset includes 107 speakers.
    \item Age: According to VCTK, speakers range in age from 18 to 38, with several intermediate values not present. These age values were treated as categorical labels rather than continuous variables, resulting in 15 unique target labels for classification.
    \item Gender: The ASVspoof 2019 LA dataset features both male and female speakers. Gender was treated as a binary classification label. There are 46 male and 61 female speakers in the datasets.
    \item Accent: VCTK includes speakers from 44 regions of the United Kingdom and other English-speaking countries. Their accents are available and were categorized as 12 classification labels. They include English, Welsh, and American, to name a few.
\end{itemize}
The other type of meta traits is about spoofing attacks used for ASVspoof 2019 LA. We derived the information from \cite{asvspoof2019} and assembled it as with two categories.
\begin{itemize}
    \item Attack ID: As later detailed in Section \ref{secsec:datasets}, we acquired the evaluation set of the original ASVspoof 2019 LA as the source dataset for re-partitioning. It contains 13 spoofing attacks which have not been seen by either pre-trained ASV or CM modules. We acquired their IDs (A07-A19) as the classification labels.
    \item Type of attack: This is about the backbone system that has been used for constructing the spoofing systems. ASVspoof 2019 LA scenario was built via mainly two types of synthesis systems widely-known, and we broadly categorized them into two types: TTS and VC\footnote{For attack A13-A15 where TTS is followed by VC\cite{asvspoof2019}, we broadly regard them as ``TTS" in this study.}.
\end{itemize}
Note that in the evaluation set, there are also certain amount of bonafide audios which, thus, are not applicable to any of the above categories. We simply assigned ``bonafide" as a separate class label for each trait, for those audios.

\subsection{Acoustic probing traits (for regression)}
\label{secsec:physicals}
Different from meta traits readily available in the VCTK corpus metadata, the acoustic properties need to be measured. This can be achieved using established toolkits or estimation algorithms, and used as target values for regression. We measure the following traits:
\begin{itemize}
    \item Fundamental frequency (F0): The F0 value encoded in the ASVspoof involved speakers are mostly flat and less-varied than other in-the-wild datasets such as VoxCeleb. We extracted the F0 value of the speaker in the input audio using \emph{parselmouth}\footnote{\url{https://github.com/YannickJadoul/Parselmouth/}}, where the F0 values are computed at frame-level and averaged across the frames for each input audio sample.
    \item Speaking rate: the original bonafide data, sourced form VCTK, preserves clean acoustic condition, which makes speaking rate of the speakers being able to be measured in a valid way, due to less interference on the encoding of relevant features into the embeddings. However, this might not always be the case for spoofed audio. Here, we defined the speaking rate simply as \emph{words-per-second}.
    \item Duration: The duration length of the audio can also be preserved as part of the information. We simply treated such length in seconds as scalar value, to perform the regression task.
    \item Signal-to-noise ratio (SNR): SNR presents level of a desired signal relative to the level of background noise, in the form of a single value for each audio. We used LibROSA \cite{mcfee2015librosa} to estimate the signal and noise power and compute SNR. SNR is typically expressed in decibels (dB). 
\end{itemize}

\section{Data and Backbone Setups}
\label{sec:data_setups}

\subsection{Data partitioning}
While ASVspoof 2019 LA provides a standard experimental protocol and data partitioning, our focus is on the explainability of embeddings. We selected the evaluation set as our source dataset for the experiment. We created three schemes of the dataset when doing the 90-10 split, to suit different tasks.
\begin{itemize}
    \item \texttt{T01}: This is for classification of target speaker ID. This variant acquired only the bonafide part of the ASVspoof 2019 LA, where we ensured that all speaker-related and acoustic traits have been covered, including the speaker IDs, so the number of speakers in both sets were same.
    \item \texttt{T02}: This is for classification of speaker-related meta and acoustic traits listed in Section \ref{sec:counter_facts}, apart from target speaker ID.  ame as \texttt{T01}, we acquired only the bonafide part of the original evaluation set. But here, the splitting was based on speaker IDs, which means there was no overlap on speakers between training and evaluation partitions.
    \item \texttt{T03}: This is for classification of the two spoof-related meta traits. This variant acquired the whole original evaluation part of the ASVspoof 2019 LA, including bonafide and spoofed ones. We randomly partitioned it for training and evaluation, ensuring all the variables in each spoof-related trait has been covered. 
\end{itemize}

\subsection{Backbone Models: ECAPA-TDNN \& AASIST}
For ASV and CM embeddings, we acquired state-of-the-art DNN-based pre-trained models that served as backbone models for the SASV 2022 challenge \cite{sasv2022}. For ASV embeddings, we used ECAPA-TDNN \cite{ecapa_tdnn2020}, a speaker encoder trained on VoxCeleb2, which generates speaker embeddings. We extracted these embeddings from the first fully-connected layer following the pooling layer. For CM embeddings, we employed an AASIST \cite{aasist2022} trained on the ASVspoof 2019 LA training partition, which demonstrated strong CM performance \cite{asvspoof2019}. The CM embeddings were extracted from the fully-connected layer preceding the output layer. Both embedding extraction methods align with the practices established in the SASV 2022 baseline system \cite{sasv_personal_odyssey2022}. The dimension of the extracted ASV and CM embeddings were 192 and 160, respectively.

\subsection{Evaluation metric}
\label{secsec:evalaution}
For the empirical evaluation of our models, we employ distinct metrics for classification and regression tasks.
For classification, we simply compute the percentage of the correctly classified audio samples with respect to the trait and the total number of samples. 
For regression, we utilized the $R^2$ value, which quantifies the proportion of variance in the dependent variable that is predictable from the independent variables \cite{devore2011probability}. This metric is particularly useful for assessing the model on capturing the underlying trait variations. Note that while the value offers insights into model performance, it may not fully capture the effectiveness of the model in scenarios such as SNR estimation, due to expected less variations in the input audio. Identifying such case from the model performance is part of the objective. The statistical significance of the classification accuracy and $R^2$ values were measured by computing confidence interval \footnote{\url{https://github.com/luferrer/ConfidenceIntervals}} and p-value \cite{statistical_ml_r2013}, correspondingly.
We also performed ablation study in Section \ref{secsec:ablation}, to further the findings in the performance statistics.

\begin{figure}[t]
    \centering
    \includegraphics[width=0.9\linewidth]{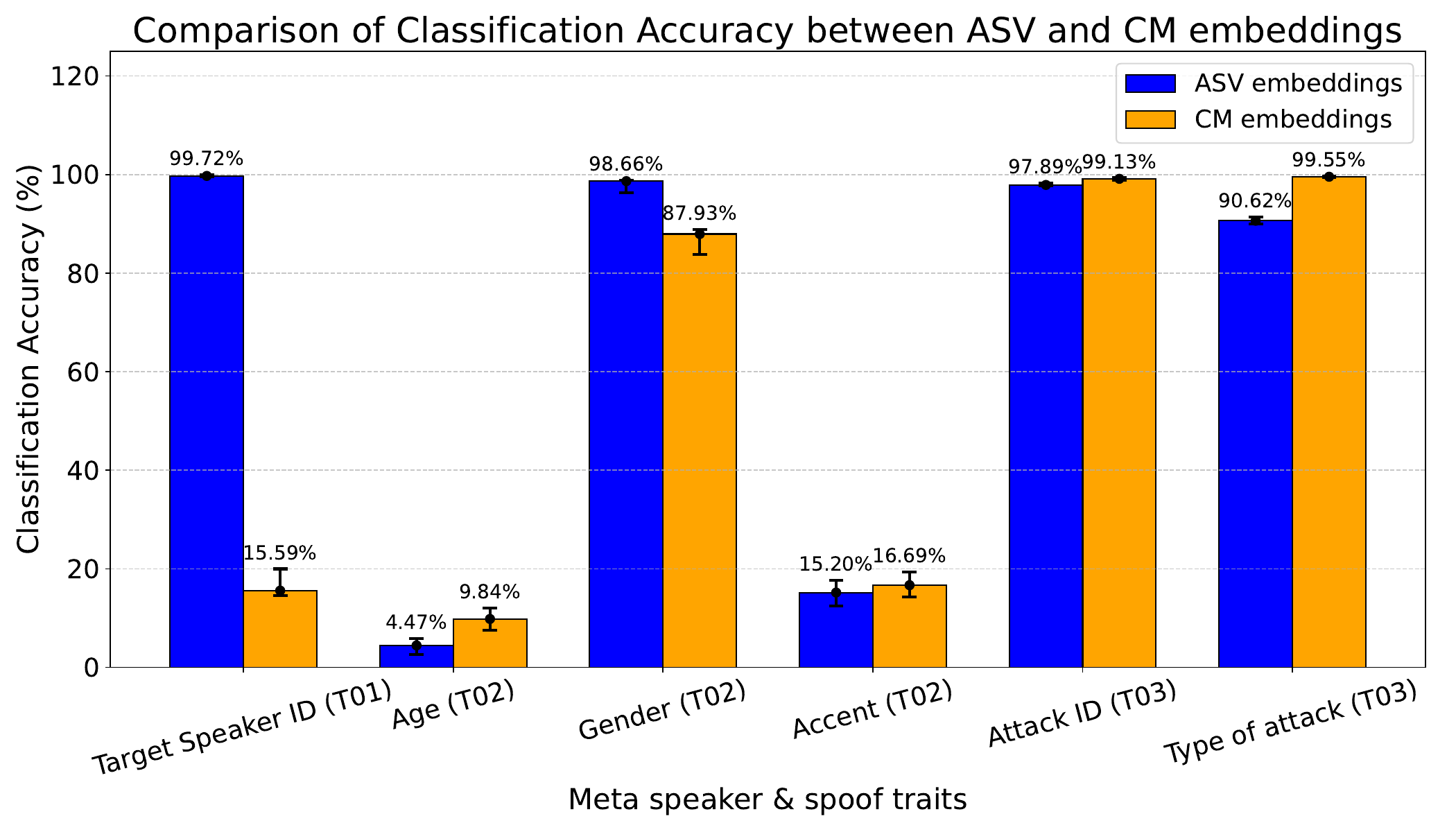}
    \caption{The classification accuracy of MLP trained using ASV and CM embeddings on various speaker and spoof-related traits available in the metadata. The brackets in the x-axis indicate their corresponding setup. The black error bars present the statistical significance of the accuracy by confidence interval measurement.}
    \label{fig:class_acc}
\vspace{-0.3cm}
\end{figure}

\begin{figure}[t]
    \centering
    \includegraphics[width=0.9\linewidth]{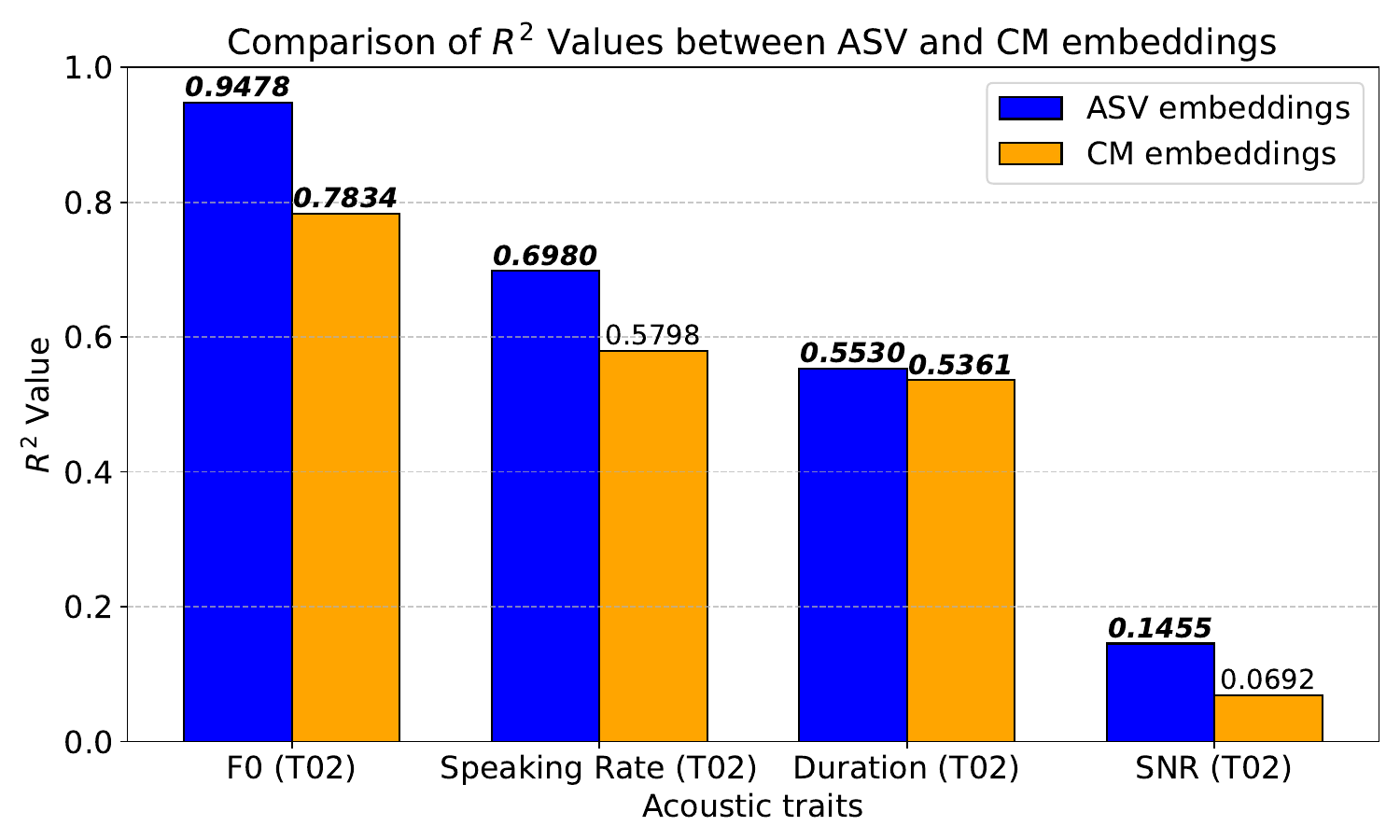}
    \caption{The $R^{2}$ values of MLP trained using ASV and CM embeddings on acoustic speaker traits. The brackets in the x-axis indicate their corresponding setup. The bold italic values indicate the results that hold p-value less than 0.01.}
    \label{fig:reg_acc}
\vspace{-0.3cm}
\end{figure}

\section{Results}
\label{sec:results}

\subsection{Empirical results}
\textbf{Gender preservation is observed from CM embeddings}. The performance statistics of classification and regression tasks for ASV and CM embeddings have been illustrated in Fig. \ref{fig:class_acc} and Fig. \ref{fig:reg_acc}, respectively. Let us first look at the classification performance. Not surprisingly, ASV embeddings performed well ($>$90\% evaluation accuracy) on target speaker identification and gender classification. Such finding demonstrates the effectiveness of ASV embeddings in leveraging meta speaker traits for speaker recognition, partially aligning with earlier findings \cite{probe_xvectors}. On the other hand, CM embeddings exhibited relatively low classification accuracy in speaker identification, suggesting that they often reduce the impact of speaker information. Meanwhile, gender has been moderately preserved in both embeddings. Some traits such as accent and age were poorly captured by both embeddings. However, this might due to VCTK containing native English speakers with limited age variation. Further dataset-level expansion might be necessary as future work.

\textbf{Both embeddings can predict the attack-related labels well}. As expected, CM embeddings effectively predict the type of attack and attack IDs, as evidenced by the high prediction accuracy shown in Fig. \ref{fig:class_acc}. Notably, ASV embeddings also achieve commendable performance on these traits, suggesting their potential in detecting spoofing attacks, consistent with findings in \cite{sasv_asv2024}. However, further studies are needed to explore and adopt ASV embeddings for contemporary speech generation techniques.

\textbf{CM embeddings retain a significant amount of acoustic and perceptual information}. Regarding regression performance, we observed that while ASV embeddings still outperform CM embeddings, the performance gap narrows for certain traits, with both embedding sets showing moderate performance on detecting data correlation. Both the sets achieved their highest $R^2$ for the F0, highlighting its potential usefulness. Neither type of embeddings produced decent predictions for SNR, which is expected given the relatively clean and uniform acoustic conditions of the ASVspoof 2019 LA dataset. For both speaking rate and audio duration, the prediction performance of the two embedding sets is moderate, with a smaller performance gap compared to the other two traits. Considering the impact of speaker variability on speaker recognition \cite{speaker_augmentation2024}, it would be valuable to explore how such variation affects CM performance, as discussed in Section \ref{secsecsec:speaking_rate_vary}.

\subsection{Ablation study}
\label{secsec:ablation}
The results illustrated in the previous subsection shows that CM embeddings preserve certain trait information. However, whether it has been effectively acquired for spoofing detection remains the question. While we believe that such topic remains broadly for future work, this section presents our primary attempts on dissecting the relationship between the traits and the spoofing detectors. We use the AASIST CM that have been acquired in this study as the case model, investigating two related questions from the empirical results.

\begin{figure}[t]
    \centering
    \includegraphics[width=0.9\linewidth]{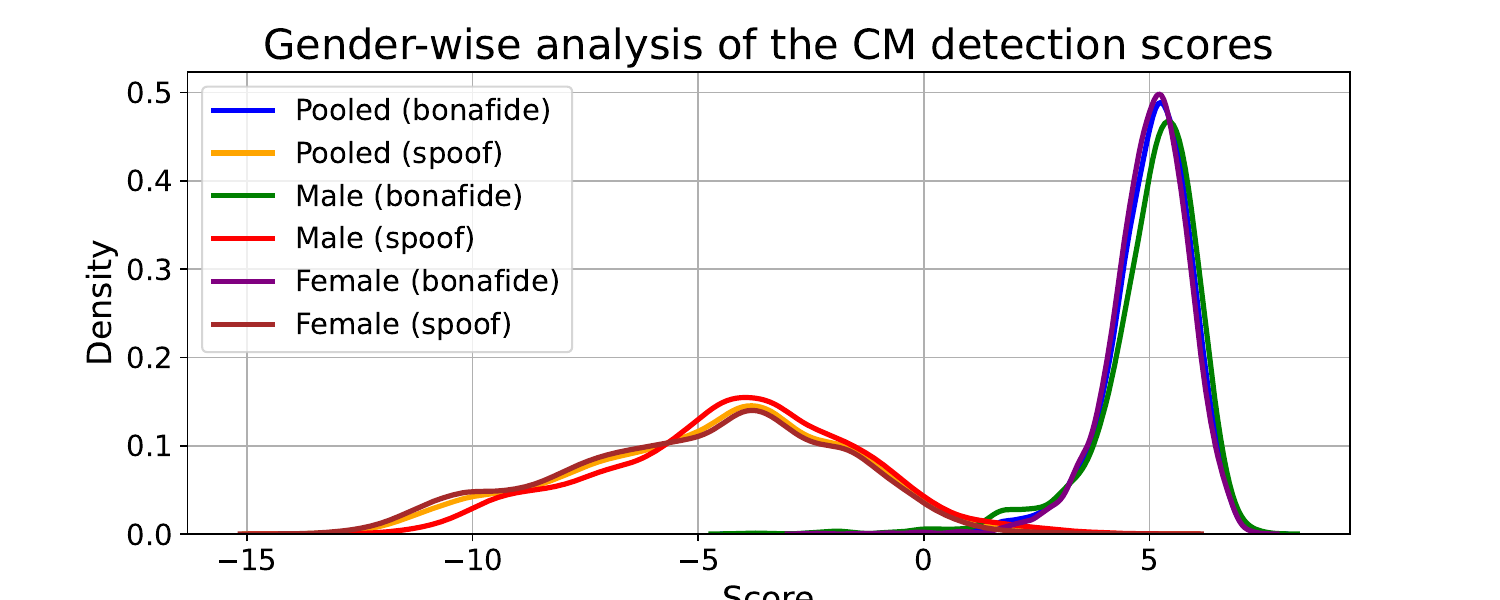}
    \caption{Score distribution of AASIST CM with respect to the gender and bonafide/spoofed cases.}
    \label{fig:gender_scores}
\end{figure}

\begin{figure}[t]
\centering
\subfloat[Training, encoder output\label{fig7:train_encoder}]{
\includegraphics[width=0.48\linewidth]{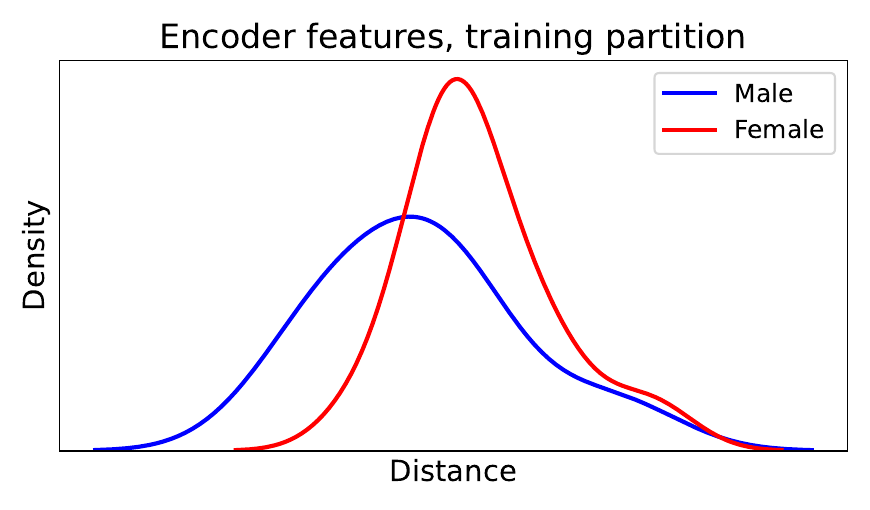} 
}
\subfloat[Evaluation, encoder output\label{fig7:eval_encoder}]{
\includegraphics[width=0.48\linewidth]{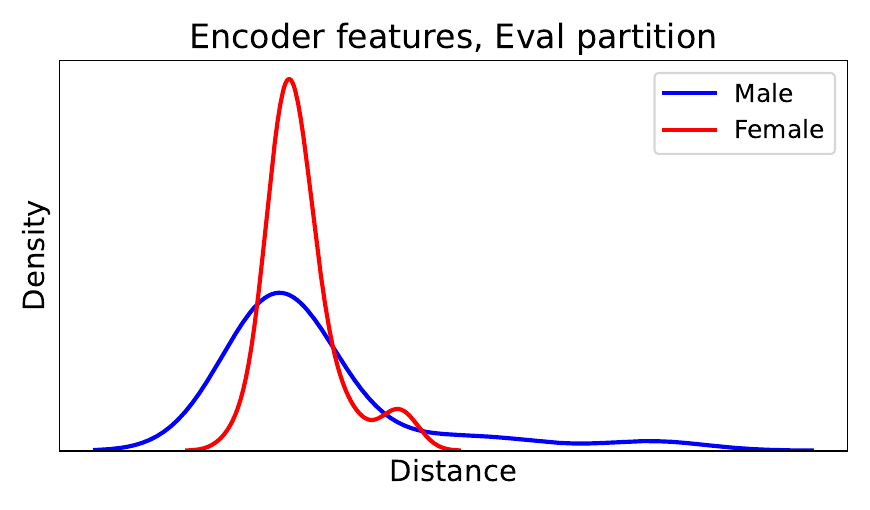} 
} \par
\subfloat[Training, embeddings\label{fig7:train_seg}]{
\includegraphics[width=0.48\linewidth]{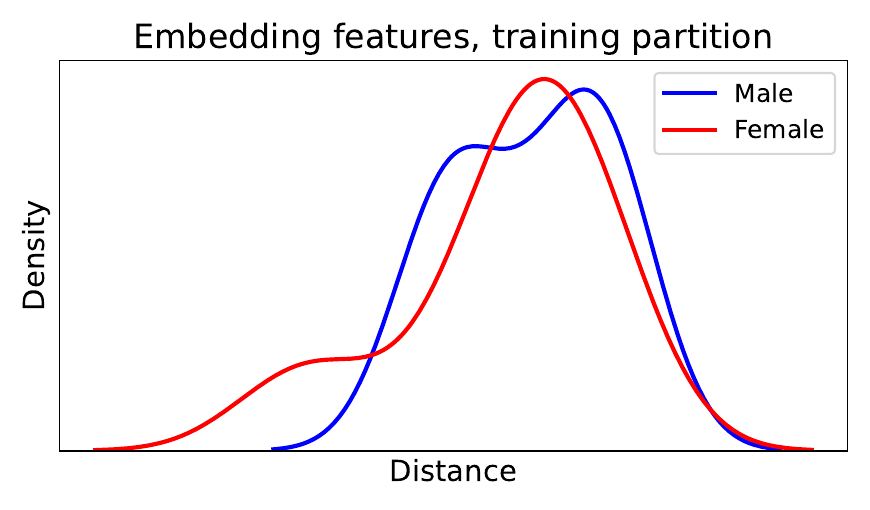} 
} 
\subfloat[Evaluation, embeddings\label{fig7:eval_seg}]{
\includegraphics[width=0.48\linewidth]{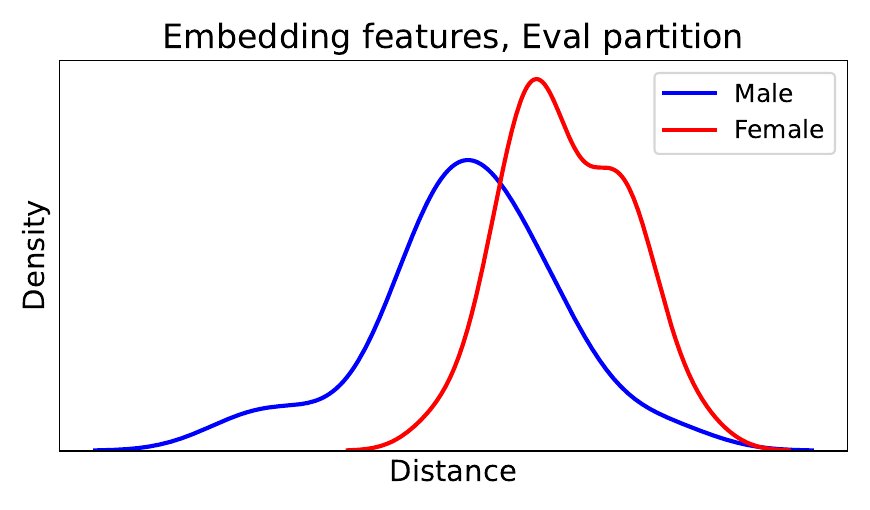} 
}
\caption{Gender-wise distance measurement between bonafide and spoof representations, aggregated per speaker, for the encoder output and embeddings from AASIST CM.}
\label{fig:gender_wise}
\vspace{-0.3cm}
\end{figure}

\subsubsection{Gender}
To investigate whether the gender information preserved in CM embeddings influences CM scoring behavior, we analyzed the CM scores of AASIST on the evaluation set of ASVspoof 2019 LA, segregating the scores by gender (male and female). The resulting score distributions, shown in Fig. \ref{fig:gender_scores}, reveal no significant difference between male and female scores compared to the pooled performance. This suggests that although gender information is present in the CM embeddings, it does not appear to influence the model's ability to distinguish between real and spoofed speakers.

Given this observation, we hypothesized that the CM might normalize gender information, enabling the embeddings to perform spoof detection in a gender-invariant manner. To explore this, we extracted representations from two parts of the AASIST CM \cite{aasist2022}: 1) The RawNet2 frontend encoder \cite{rawnet2_spoofing2021}, which outputs frame-level spectrograms, and 2) The layer preceding the decision-making layer, which provides the CM embeddings used in this study. Our methodology involved visualizing the pairwise representation distance between bonafide and spoofed audio samples within each gender subset. For each bonafide audio, we averaged the distance values between it and the spoofed audios to obtain a single paired value. For the encoder outputs, we employed the Itakura-Saito distance \cite{itakura_saito_distance}, a commonly used measure for spectral densities:
\begin{align}
D_{IS}(P, Q) &= \sum_{i=1}^{N} \left( \frac{P_i}{Q_i} - \log \frac{P_i}{Q_i} - 1 \right)
\end{align}
where $P_{i}$ and $Q_{i}$ are spectral representations of the $i$-th frame of $N$ frames. Smaller distance values indicate greater similarity between spectra, while larger ones reflect perceptual mismatch. The consistency of number of frames were achieved by chunking 4-second audio. 
For the embeddings, we used cosine similarity \cite{cosine_similarity} as the distance function.

The pairwise distance calculations were conducted for both the training and evaluation partitions, with resulting score distributions shown in Fig. \ref{fig:gender_wise}. Notably, when comparing the encoder output with the embedding representations, the distance score distributions between male and female samples converge with more overlaps as moving from the encoder output to the embeddings. This supports the hypothesis that the CM model leverages gender information to achieve gender-invariant spoof detection, effectively neutralizing gender-related bias in its decision-making process.

\begin{figure}[t]
    \centering
    \includegraphics[width=0.85\linewidth]{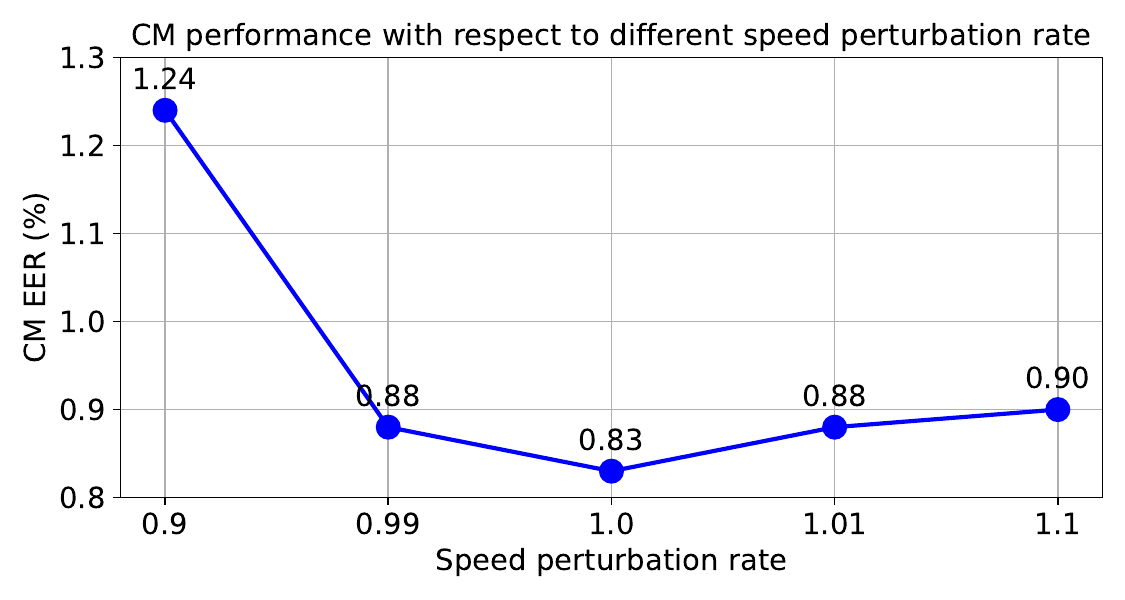}
    \caption{CM performance in EER(\%), with repsect to different speed perturbation rate. The perturbation rate being 1.0 is the baseline system, and the CM EER is same as in \cite{aasist2022}.}
    \label{fig:speaking_rate_tuning}
\vspace{-0.3cm}
\end{figure}

\subsubsection{Speaker rate \& duration fine-tuning}
\label{secsecsec:speaking_rate_vary}
We aim to investigate whether perturbing the speaking rate of speech in the evaluation set affect CM performance. We hypothesize that minor variations in speaking rate and duration have a minimal impact on CM performance, suggesting that the CM not only retains but also leverages these traits for robust spoofing detection, effectively accommodating corresponding variations.

To validate this, we applied speed perturbation to the dataset using five different rates, as shown on the x-axis of Fig. \ref{fig:speaking_rate_tuning}. The resulting CM performance is also presented in Fig. \ref{fig:speaking_rate_tuning}. While larger perturbation rates significantly degrade CM performance, these rates likely alter key speaker characteristics originally preserved in the CM representations, such as F0. Thus, the observed performance decline highlights the capability of CM to acquire and utilize speaking rate and duration, confirming their role in spoofing detection.

\section{Conclusion}
\label{sec:conclusion}
In summary, this study has investigated the explainability of embedding representations from modern spoofing detection systems. Using state-of-the-art CM as an example, we performed a probing analysis case by re-partitioning the original ASVspoof 2019 LA dataset, extracting explainable information in the form of various trait labels, and applying simple neural network models for classification and regression. Our findings indicate that, unlike embeddings commonly extracted from speaker encoders, which retain rich speaker-related information, modern neural-based spoofing detectors tend to discard most speaker-related meta-traits except for gender, while moderately preserving spoof-related meta and acoustic traits. Ablation study on gender and speaking rate reveals that spoofing detectors retain this information primarily to enhance the robustness and invariance of CM to these factors. Future work could focus on improving CM performance and robustness by leveraging the preserved information and effectively integrating the missing information into CM embeddings. This primary research also points to an intriguing direction for integrating ASV and CM systems, focusing on preserving information.

\bibliographystyle{IEEEtran}
\bibliography{strings}

\end{document}